\documentstyle[aps,twocolumn,epsf,prl]{revtex}
\begin{document}
\title{Nonclassicality and the concept of local constraints on the
photon number distribution}
\author{R. Simon~\cite{email1}}
\address{Institute of Mathematical Sciences, CIT Campus,
Madras 600 113\\ India}
\author{Mary Selvadoray}
\address{Institute of Mathematical Sciences, CIT Campus,
Madras 600 113\\ India}
\author{Arvind~\cite{email2}}
\address{Department of Physics, 
Indian Institute of Science, Bangalore 560 012\\ India}
\author{N. Mukunda}
\address{Department of Physics and Center for Theoretical Studies,\\
Indian Institute of Science, Bangalore 560 012 \\India}
\maketitle
%\pacs{42.50.~Dv, 03.65.~Fd, 42.50.~Ar, 42.50.~Lc} 
\begin{abstract}
We exploit results from the classical Stieltjes moment problem to bring
out the totality of all the information regarding phase insensitive
nonclassicality of a state as captured by the photon number distribution
$p_n$. Central to our approach is the realization that $n !\, p_n$ 
constitutes the sequence of moments of a (quasi)~probability
distribution, notwithstanding the fact that $p_n$ can by itself be
regarded as a probability distribution. This leads to classicality
restrictions on $p_n$ that are local in $n$ involving $p_n$'s for only
a small number of consecutive $n$'s, enabling a critical examination of
the conjecture that oscillation in $p_n$ is a signature of 
nonclassicality.
\end{abstract}
\vspace*{12pt}
Nonclassical states of the radiation field continue to receive much
attention. These are states for which the $P$-distribution $\varphi(z)$ 
is not a true probability. Prominent among the quantitative
characteristics of nonclassicality are squeezing and sub-Poissonian
statistics. While these involve the lower order moments of $\varphi(z)$,
there have also emerged criteria involving the higher order moments.
Among these we may note the higher order squeezing criteria of Hong and
Mandel~\cite{mand}, the related amplitude squared squeezing introduced
by Hillery~\cite{hill}, and the generalization of the Mandel Q-parameter
achieved by Agarwal and Tara~\cite{tara}.

There has also emerged a qualitatively different kind of criterion for
nonclassicality. While $p_n = \langle n | \hat{\rho} | n \rangle$, which
represents the probability of there being $n$ photons in the state
specified by the density operator $\hat\rho$, is a smooth function of
$n$ for classical states like the coherent states and the thermal
states, it is an oscillating function of $n$ for nonclassical states
like the squeezed states, as was exposed in the seminal work of Schleich
and Wheeler~\cite{osc} on interference in phase space. Oscillation in
$p_n$ has since then been taken as a signature of 
nonclassicality~\cite{osc2,dutta}. Indeed, these oscillations have
come to be known as {\it nonclassical oscillations}~\cite{caves}. It
should, however, be noted that this oscillation criterion for
nonclassicality, though insightful, has not been derived from basic
principles and hence enjoys only the status of a conjecture. Its
principal virtue lies in the fact that it is $local$ in $n$, in 
contradistinction to the criteria involving the Mandel Q-parameter
or its generalizations; the latter are expressed in terms of the moments
of $p_n$, and hence are {\it global} in $n$.

For a radiation mode described by annihilation and creation operators
$\hat{a}$, $\hat{a} ^{\dagger}$ measurements of operators which are 
functions of $\hat{a} ^{\dagger} \hat{a}$ (the so called phase
insensitive operators) do not depend on all the details of
$\varphi (z)$, but are fully determined by the angle averaged radial
``marginal'' distribution ${\cal P} (I)$, derived from $\varphi(z)$ by
writing $z = I^{1/2}e^{i\theta}$ and averaging over $\theta$:
%01
\begin{eqnarray}
{\cal P} (I) = \frac{1}{2\pi}\int^{2\pi}_{0} d\theta
\varphi (I^{1/2}e^{i\theta})\,.
\label{1.eq1}
\end{eqnarray}
In particular we have
%02
\begin{eqnarray}
p_n~=~
\int^{\infty}_0 dI \tilde{\cal P} (I) \frac{I^n}{n!}\,, \qquad
\tilde{\cal P} (I) = {\cal P} (I) e ^{ - I}\,,
\label{1.eq2}
\end{eqnarray}
where $n = 0, 1, 2, \ldots$ The above relation is invertible. That is,
the sequence $\{ p_n \}$ represents
${\cal P} (I)$ faithfully.

For a given state, it may happen that $\varphi(z)$ is not a true
probability, but the phase averaged ${\cal P} (I)$ is a bonafide
probability distribution. Such states (the Yurke-Stoler
state~\cite{yurk} is an example) are said to exhibit phase sensitive
nonclassicality. On the other hand the nonclassicality of the state may
be such that it survives the process of phase averaging involved 
in~(\ref{1.eq1}), thereby rendering ${\cal P} (I)$ itself a
quasiprobability rather than a true probability. Then we talk
of~(the stronger) phase insensitive nonclassicality. Clearly, any state
with sub-Poissonian statistics is nonclassical of the phase insensitive
type.

The purpose of this Letter is to exhibit the totality of all the
information regarding nonclassicality of a state as captured by the 
photon number distribution sequence $\{p_n\}$ or, equivalently, by 
${\cal P} (I)$. Since we work at the level of $\{p_n\}$, and not 
$\varphi (z)$, only states with phase insensitive nonclassicality will
be said to be ``nonclassical''. All other states will be termed as 
``classical'', for brevity.

{\it The key to our approach is an appreciation of the fact that
$\{p_n\}$ is essentially the {\it moment sequence} of a
(quasi)~probability distribution, notwithstanding the fact that
$p_n \ge 0$ and $\sum p_n = 1$, and hence $\{p_n\}$ can legitimately be
viewed as a probability distribution over the discrete variable $n$}.
This departure from tradition leads us to derive constraints on $p_n$
which are local in $n$ involving $p_n$'s for only a small number of
consecutive $n$'s, and enables us to critically examine the oscillation
criterion in a direct manner. Necessary and sufficient conditions for 
absence of~(phase insensitive) nonclassicality in a state are presented,
not only in terms of the sequence $\{p_n\}$ but also in the 
(dual)~traditional approach involving the factorial moments of
$\{p_n\}$.

\noindent
{\it Local constraints on classical $\{ p_n \}$.}---
It turns out to be convenient to define a sequence $\{q_n\}$ in the
place of $\{p_n\}$ through $ q_n = n! p_n$, for $n = 0, 1, 2, \ldots$ It
follows from~(\ref{1.eq2}) that $\{q_n\}$ is simply the moment sequence
of the distribution $\tilde{\cal P} (I) = {\cal P} (I) e ^{- I}$:
%03
\begin{eqnarray}
q_n~=~\int^{\infty}_{0} dI \tilde{\cal P} (I) I^{n} \equiv <I^n>_ 
{\tilde{\cal P}}\,.
\label{1.eq3}
\end{eqnarray}

Now suppose we are given a classical state so that 
$\tilde{\cal P} (I)\geq 0$, for $0\leq I < \infty$, and consider the
polynomial $f(I) = I^{n}(I - x)^{2}$. Since $f(I)$ is manifestly
nonnegative for any real value of the parameter $x$, nonnegativity of
$\tilde{\cal P} (I)$ implies, through~(\ref{1.eq3}),
%04
\begin{eqnarray}
<f (I)>_{\tilde{\cal P}}
& = &
\langle x ^2 I ^n - 2 x I ^{n + 1} + I ^{n + 2}
\rangle _{\tilde{\cal P}} \nonumber \\
& = & x^2 q_n - 2 x q_{n+1} + q_{n+2}~\geq~0\,,
\label{1.eq4}
\end{eqnarray}
for all real $x$. That is, $q _n\,, q _{n +2} \ge 0$ and
%05
\begin{equation}
q_n q_{n+2}~\geq~q^{2}_{n+1}\,, \qquad n = 0, 1, 2,\ldots
\label{1.eq5}
\end{equation}
Written in terms of $\{p_n\}$, the above condition reads
\begin{eqnarray}
p_n p_{n+2}~\geq~(\frac{n+1}{n+2}) p^{2}_{n+1}~.
\label{1.eq6}
\end{eqnarray}
These are our {\it local conditions} to be necessarily satisfied by the
photon distribution $\{p_n\}$ of any classical state.

Several interesting conclusions can be drawn from these conditions which
are local in $n$, and are saturated for every value of $n$ by any
Poissonian distribution. Suppose that we are given a state for which
$p_{n_0}=0$ (and hence $q_{n_0}=0$) for some integer $n_0 \geq 0$, and
assume that the state is classical. The choice $n=n_0$ in the local
condition~(\ref{1.eq5}) implies that $q_{n_0 +1}=0$. Similarly the
choice $n + 2 = n_0$ implies $q _{n_0 -1} = 0$. Continuing this process
we find that for a classical state either $p_n$ is nonzero for every
values of $n$, or $p_n=0$ for all $n>0$. In other words, 
{\it a classical
state other than the vacuum state, cannot be orthogonal to any Fock
state}. To appreciate the significance of this conclusion, consider the
state 
%07
\begin{equation}
\hat{\rho} = N \hat{a} ^{\dagger m}\hat\rho_{0} \hat{a} ^{m}~,
%\label{1.eq7}
\end{equation}
where $\hat\rho_{0}$ is an arbitrary density operator, and $N$ is a
normalization constant. We can call it the ``photon added''
$\hat\rho_{0}$, for it includes the photon added coherent 
state~\cite{gsa} and the photon added thermal state~\cite{tara,jones}
as special cases. Since $p_n = <n|\hat{\rho}|n> = 0$ for $n<m$, we
conclude that $\hat{\rho}$ is nonclassical for every $m > 0$. Thus, we
have established the following result: 
{\it all photon added states, pure or mixed, are nonclassical}.

There has been remarkable progress recently in quantum state
reconstruction using techniques of optical homodyne 
tomography~\cite{breitenbach}. Thus, it is now possible to `map out' the
Wigner distribution of a state using the inverse Radon transform, or
reconstruct the density matrix in the Fock basis using a set of pattern
functions. Schiller~{\it et al}~\cite{schiller} report such a
reconstructed $\hat{\rho}$ upto $n = 6$, with $q _n = 0.44$, $0.07$,
$0.26$, $0.30$, $1.44$, $3.60$, $28.80$. The local
conditions~(\ref{1.eq5}) are violated~(for instance, 
$0.07 \times 0.30 \le (0.26) ^2$). Thus, the squeezed vacuum of
Schiller~{\it et al} turns out to be a nonclassical state of the phase
insensitive type. That is, the nonclassicality of their state survives 
phase averaging, notwithstanding the fact that consideration of the
Mandel $Q$-parameter will do no more than to simply indicate that this
state is ``strongly super-Poisonian'' as noted by the authors.

We now turn to the oscillation criterion for nonclassicality. There
exist classical states for which the photon

\noindent
distribution $\{p_n\}$ is 
an oscillatory function of $n$. We may call these {\it classical
oscillations}.  Figure~1 shows such classical oscillations for an
incoherent mixture of suitably chosen coherent states:
%07
\begin{equation}
\hat{\rho} = \sum \lambda _j 
\left| \alpha _j \rangle \langle \alpha _j \right|\,, \qquad
\sum \lambda _j = 1\,.
\label{1.eq7}
\end{equation}
\vspace*{-2cm}
\begin{figure}[htb]
\begin{center}
\epsfxsize=8cm
\epsfbox{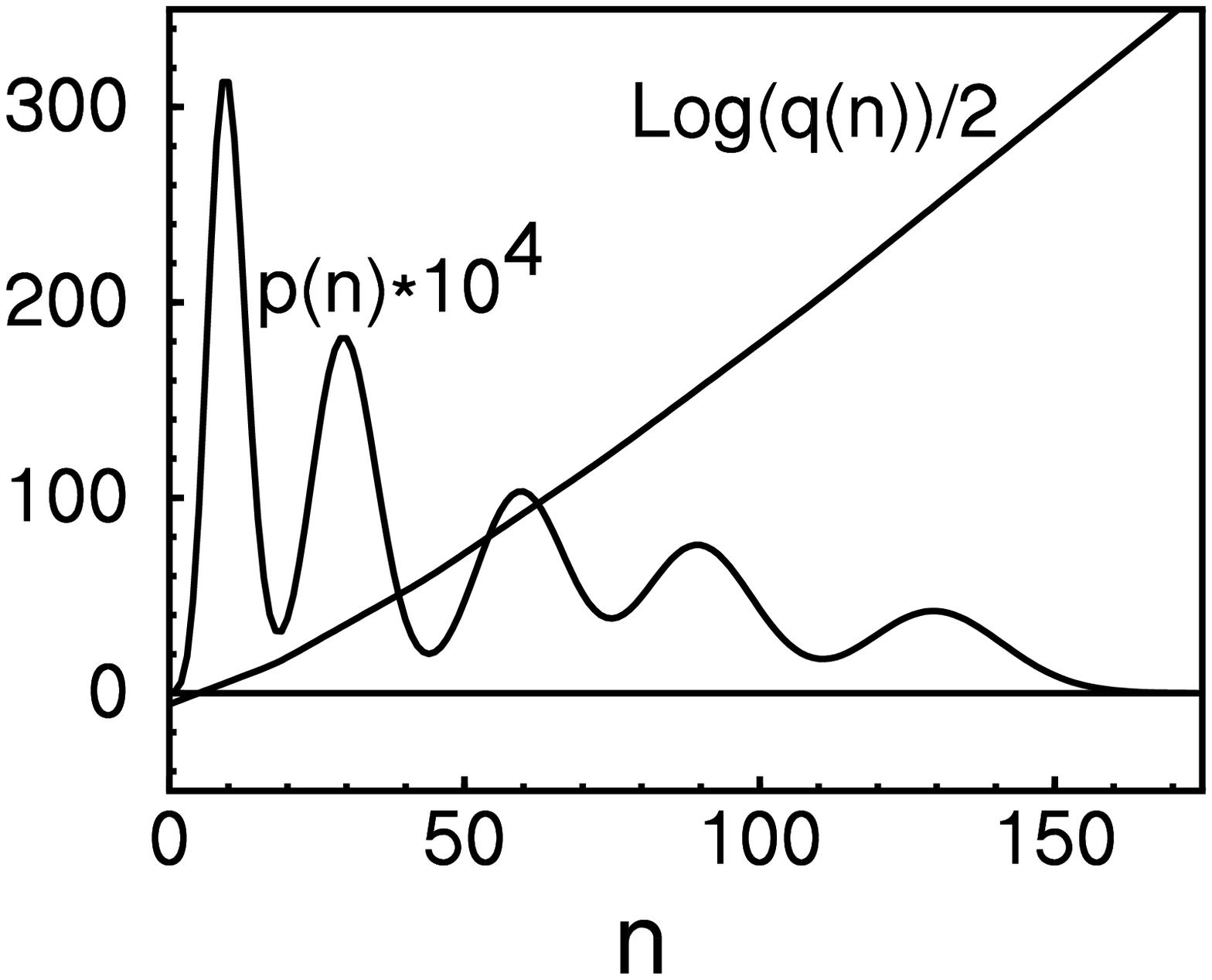}
\caption{
Showing classical oscillations in the photon distribution for an
incoherent mixture of five coherent states with $\lambda _j = 0.25$,
$0.25$, $0.2$, $0.18$, $0.12$ and corresponding $| \alpha _j | ^2 = 10$,
$30$, $60$, $90$, $130$. Here $p (n)$ and $q (n)$ stand, respectively,
for $p_n$ and $q_n$ of the text. Note that $q_n$ exhibits no oscillation
for this classical state}.
\label{Fig 1.}
\end{center}
\end{figure}
On the other hand the photon added thermal~(or coherent) state has a
$\{ p_n \}$ with no oscillation; we have nevertheless seen that it is a
nonclassical state. This, however, should not tempt one to simply
dismiss the
oscillation criterion as being neither sufficient nor necessary for
nonclassicality; for, as already noted, the virtue of this criterion
does not reside in its exactitude, but rather in its distinction of
being local in $n$. Thus, it is highly desirable  to amend it suitably
but without sacrificing its local character. This is easily achieved
through our local conditions. To see this, note that~(\ref{1.eq5})
implies that $\{ q_n \}$ for a classical state cannot have a local
maximum (if it had, the inequality will be violated by allowing $n + 1$
to correspond to the local maximum), and hence cannot exhibit any
oscillation. Thus we arrive at the desired modification: 
{\it oscillation in $\{ q_n \}$ is a sufficient condition for
nonclassicality}. It is $\{q_n\}$, and not $\{p_n\}$, that is the key 
to the correct oscillation criterion. Indeed, $\{p_n\}$ can oscillate
for a classical state with amplitude limited by the extent permitted by
the $(n+1)/(n+2)$ factor in~(\ref{1.eq6}). Even period-two classical
oscillations are allowed, as can demonstrated using classical states
of the type~(\ref{1.eq7}).

Finally, we apply our local conditions to a class of states obtained as
superposition of two coherent states:
%09
\begin{eqnarray}
|\Psi> = N[|z_0>+e^{i\theta}|-z_0>]  , \label{psi}
\end{eqnarray}
where $\theta$ is the relative phase (in the Pancharatnam~\cite{panch}
sense) between the two components of the superposition, and $N$ is the
normalization constant. We have
%10
\begin{eqnarray}
\frac{q_n q_{n+2}}{q^{2}_{n+1}} = \frac{(1+(-1)^{n} \cos \theta)^{2}}
{(1+(-1)^{n+1} \cos \theta)^{2}}\,. \nonumber \\
\end{eqnarray}
It is clear that our local conditions~(\ref{1.eq5}) are violated by
$\left| \Psi \right. \rangle$ for all values of
$\theta \neq \pm \pi /2$; by odd values of $n$ for $-\pi /2 < \theta
< \pi /2$, and by even values of $n$ for the range
$-3 \pi /2 < \theta < -\pi /2$. It is well known~\cite{pure} that 
coherent states are the only pure states for which the
${\cal P}$-distribution $\varphi(z)$ is a true probability. Further, the
Yurke-Stoler~\cite{yurk} states, which correspond to 
$\theta = \pm \pi /2$, have Poissonian $\{p_n\}$ and hence possess only
phase sensitive nonclassicality. Thus, what is striking about the above
analysis is the inference that for all values of
$\theta \neq \pm \pi /2$ the superposition state has 
{\it phase insensitive} nonclassicality, and that it is exposed by our
lowest order local conditions!

\noindent
{\it Necessary and sufficient condition for nonclassicality.}---
We showed that positivity of $\tilde{\cal P} (I)$ implies the local
conditions~(\ref{1.eq5}) on its moment sequence $\{q_n\}$. We now
exploit results from the classical problem of moments to exhibit the
necessary and sufficient conditions on $\{q_n\}$ in order that the
associated state $\hat \rho$ is classical.

The classical moment problem, on which there exists an enormous amount
of literature~\cite{shoh}, consists of two parts: (i) to test if a
given sequence of numbers qualifies to be the sequence of moments of
some bonafide probability distribution, and (ii) to reconstruct a
probability distribution from its moment sequence. If the probability 
distribution is over the semi-infinite real line $[0,\infty)$, one calls
it the Stieltjes moment problem. The Hamburger moment problem
corresponds to the case where the probability distribution is over the
entire real line $(-\infty, \infty)$. Since, $I = | z | ^2 \ge 0$, our
problem of deriving the necessary and sufficient condition on the
moment sequence $\{q_n\}$ in order that $\tilde{\cal P} (I)$ is a true
probability distribution is indeed a Stieltjes moment problem. 

Solution of this classical problem is well known~\cite{shoh}. To exhibit
this solution, construct from the moment sequence $\{q_n\}$ two
$(N+1)$-dimensional symmetric square matrices $L^{(N)}$ and
$\tilde L^{(N)}$ as follows:
%11
\begin{eqnarray}
L^{(N)}_{mn} & = &
\left(
\begin{array}{ccccc}
q_0 & q_1 & q_2 & \cdots & q_N \\
q_1 & q_2 & q_3 & \cdots & q_{N + 1} \\
\vdots & \vdots & \vdots &  & \vdots \\
q_{N} & q_{N + 1} & q_{N + 2} & \cdots & q_{2 N} \\
\end{array}
\right)\,, \nonumber \\
\tilde L^{(N)}_{mn} & = &
\left(
\begin{array}{ccccc}
q_1 & q_2 & q_3 & \cdots & q_{N + 1} \\
q_2 & q_3 & q_4 & \cdots & q_{N + 2} \\
\vdots & \vdots & \vdots &  & \vdots \\
q_{N + 1} & q_{N + 2} & q_{N + 3} & \cdots & q_{2 N + 1} \\
\end{array}
\right)\,. 
\label{1.eq8}
\end{eqnarray}

{\it Theorem 1:}
The necessary and sufficient condition on the photon number distribution
sequence $\{q_n=n! p_n\}$ of a state $\hat{\rho}$, in order that the
associated quasiprobability distribution $\tilde{\cal P} ( I )$ is a
true probability, is that the matrices $L^{(N)}$, $\tilde L^{(N)}$ be
nonnegative:
%12
\begin{eqnarray}
L^{(N)} \geq 0\,, \quad  \tilde L^{(N)} \geq 0\,, 
\quad N=0, 1, 2, \ldots
\label{1.eq9}
\end{eqnarray}

It may be noted in passing that for the Hamburger moment problem on
the entire real line $\left( - \infty , \infty \right)$, the condition
$L ^{(N)} \ge 0$ is both necessary and sufficient. 

It is immediate to relate our local condition to the above theorem.
Nonnegativity of $L^{(N)},  \tilde L^{(N)}$ demands, as a necessary
condition, nonnegativity of the diagonal $2 \times 2$ blocks of
$L^{(N)}, \tilde L^{(N)}$. This is precisely what our local
conditions~(\ref{1.eq5}) are! It is also clear why our local
conditions~(\ref{1.eq5}) are not sufficient: positivity of
the diagonal $2 \times 2$ blocks of $L^{(N)}, \tilde L^{(N)}$ does not
capture in its entirety the positivity of $L^{(N)}$ and 
$\tilde L^{(N)}$ required in~(\ref{1.eq9}).

We can derive the next higher level of local conditions for classicality
using our necessary and sufficient conditions~(\ref{1.eq9}). Given the
sequence $\{q_n\}$, we define
%13
\begin{eqnarray}
x_n = {q_n q_{n+2}}/{q^2_{n+1}}\,, \qquad  n = 0, 1, 2, \ldots
\label{1.eq12}
\end{eqnarray}
Then our first order local conditions~(\ref{1.eq5}) involving $q_n$ for
three successive values of $n$ simply reads $x_n \geq 1$, $\forall n$,
for any classical state. The second order local condition to be
presented involves $q_n$ for five successive values of $n$ or,
equivalently, $x_n$ for three successive values of $n$.

A necessary condition for the nonnegativity of $L^{(N)}$,
$\tilde L^{(N)}$ is that their diagonal $3 \times 3$ blocks~(such a
block involves $q_n$ for five successive values of $n$) be nonnegative.
After some algebra this condition can be written in terms of the
$x_n$'s as
%14
\begin{eqnarray}
(x_n -1)(x_{n+2}-1) \geq (\frac{x_{n+1}-1}{x_{n+1}})^{2} ,
\label{1.eq13}
\end{eqnarray}
for $n = 0, 1, 2, \ldots$ These are our {\it second order local
conditions} on $\{q_n\}$ or, equivalently, on $\{p_n\}$. They involve
three successive $x_n's$ and hence five successive $p_n's$. Just like
the first order conditions, these too are only necessary conditions for 
classicality, and we can similarly derive successive higher levels of
local conditions.

To see an interesting implication of~(\ref{1.eq13}), recall that a
Poissonian distribution $\{p_n\}$, {\it i.e.} a geometric sequence
$\{q_n\}$, saturates the first order local conditions and renders
$x_n = 1$ identically. We now ask whether it is possible to have a
classical state for which $q_n q_{n+2} = q^{2}_{n+1}$ for some values 
of $n$, whereas $q_n q_{n+2} > q^{2}_{n+1}$ for other values of $n$.
Such classical states, if they exist, can be said to be {\it locally
Poissonian} at these former values of $n$.

Suppose a classical state is locally Poissonian at some $n=n_0$.
That is, $x_{n_0}=1$. Then two applications of~(\ref{1.eq13}), once with
$n_0=n$ and then with $n_0=n+2$, shows that the state will cease to be
classical unless $x_{n_0 +1}=1$ and $x_{n_0 -1}=1$. Continuing this
process we find that $x_n=1$ for all $n$. Thus, there exists no
non-Poissonian 
classical state which is locally Poissonian: {\it A classical state is
either everywhere locally Poissonian $(x_n=1$ for all $n)$ or is
everywhere locally super-Poissonian $(x_n > 1$ for all $n)$}.

In the light of this result we can now strengthen and refine our first
order condition~(\ref{1.eq5}) by adding that for a classical state these
inequalities are either saturated for all $n$, or they are strict
inequalities for all $n$.

\noindent
{\it Factorial moments.}---
We now present a dual approach to nonclassicality based on the
traditional normal ordered moments 
$\gamma _n = {\rm tr} \left(\hat{a} ^{\dagger n} \hat{a} ^{n}
\hat {\rho}\right)$.
This approach will be seen to be along the lines of Agarwal and
Tara~\cite{tara}. However, the conditions for nonclassicality that we
present are both {\it necessary} and {\it sufficient}.

Suppose we have a state $\hat {\rho}$ whose normal ordered
moments $\gamma _n$~({\it i.e.} factorial moments
$\sum _k \left( k ! \right) ^{- 1} (n + k) ! p _{n + k}$ of $p_n$) are
known.  Our problem is to find necessary and sufficient conditions on
the sequence $\{\gamma _n\}$ in order that the state $\hat {\rho}$ is
classical. Transcribing $\gamma _n$ to the representation in terms of
the ${\cal P}$-distribution $\varphi(z)$, and writing
$z=I^{1/2}e^{i\theta}$, we have 
%15
\begin{eqnarray}
\gamma _n = 
\int^{\infty}_0 dI {\cal P}(I) I^{n} = \langle I^{n}
\rangle_{\cal P}\,.
\label{1.eq14}
\end{eqnarray}
That is $\{\gamma _n\}$ is the moment sequence of ${\cal P} (I)$, in
exactly the same manner in which the sequence $\{q_n\}$ was related to
$\tilde{\cal P} (I)$. And the state $\hat{\rho}$ being classical is
equivalent to ${\cal P} ( I )$ being a true probability distribution.
Thus, we have a Stieltjes moment problem once again, with solution
parallel to the earlier one. Using the moment sequence $\{\gamma _n\}$,
form $(N+1)$-dimensional symmetric matrices $M ^{(N)} ,\tilde M ^{(N)}$
defined by
%16
\begin{eqnarray}
M_{j k} ^{(N)} = \gamma _{j + k}\,,  \qquad
\tilde M_{j k} ^{(N)} = \gamma _{j + k + 1}\,.  \qquad
\end{eqnarray}
where $j, k = 0$, $1, \ldots\,, N$ and $N = 0, 1, 2, \ldots$\\
{\it Theorem 2:} The necessary and sufficient condition that the state 
$\hat{\rho}$ with normal ordered~({\it i.e.} factorial) moment sequence
$\{\gamma _n\}$ be classical is that
%16
\begin{eqnarray}
M^{(N)} \geq 0\,, \quad \tilde M^{(N)} \geq 0\,, \quad  
N = 0, 1, 2, \ldots
\label{1.eq15}
\end{eqnarray}

It should be appreciated that theorem~2 completes the work initiated by
Agarwal and Tara~\cite{tara} by improving their necessary condition for
classicality~(they had only the condition $M^{(N)}\geq 0$) into the
necessary and sufficient condition~(\ref{1.eq15}). Thus, the constraints
on the factorial moments $\{\gamma _n\}$ arising from the requirement
$M^{(N)}\geq 0$ are the same as in their work. However the additional 
constraints on these moments arising from the positivity requirement on
$\tilde M^{(N)}$ are new: with $N=0$ we have $\gamma _1 \geq 0$, with
$N = 1$ we have $\gamma _1 \gamma _3 \geq \gamma _2 ^2$, and so on. It
should be appreciated that these conditions cannot indeed be deduced 
from $M ^{( N )} \ge 0$.

Considering diagonal $2 \times 2$ blocks of $M^{(N)}$, $\tilde M^{(N)}$
we obtain the classicality conditions 
$\gamma _k \gamma _{k+2} \geq \gamma ^2_{k+1}$, for $k = 0$,
$1$, $2, \ldots$ Clearly, these are dual to our first order local 
conditions~(\ref{1.eq5}). We may also derive conditions analogous to
our second order local conditions~(\ref{1.eq13}), and so on.

To conclude, either of the two approaches based respectively on the
moments of $\hat{\cal P} (I)$ and ${\cal P} (I)$ leads to complete
solution of the problem of~(phase insensitive) nonclassicality as coded
in the photon number distribution $\{ p_n \}$. It should however be
appreciated that $q _n$'s are well defined for every state whereas the
factorial moments $\gamma _n$ may not be finite, for $p _n$ may not
decay fast
enough as a function of $n$. For the states for which $\gamma _n$ exists
for all $n$, the two approaches are equivalent. Even then, it is
unlikely that the connection between nonclassicality and oscillation in
$p _n$ could have been so easily settled in terms of $\gamma _n$.
Finally, the first approach in terms of local conditions of $p _n$ has a
distinct advantage at least in situations where, for some reason,
$p _n$ is known not for all values of $n$. The density matrix from
Ref.~\cite{schiller} which we analysed is such an example. The point is,
even with knowledge of $p _n$ only for a finite set of values of $n$
one can now look for signatures of nonclassicality.

\end{document}